\documentclass[%
 aip,
 amsmath,amssymb,
 reprint,%
]{revtex4-1}
\usepackage{subcaption}
\usepackage{graphicx}% Include figure files
\usepackage{dcolumn}% Align table columns on decimal point
\usepackage{bm}% bold math
\usepackage{multirow}
\usepackage{subcaption}
\usepackage[utf8]{inputenc}
\usepackage[T1]{fontenc}
\usepackage{mathptmx}
\usepackage{etoolbox}

\makeatletter
\def\@email#1#2{%
 \endgroup
 \patchcmd{\titleblock@produce}
  {\frontmatter@RRAPformat}
  {\frontmatter@RRAPformat{\produce@RRAP{*#1\href{mailto:#2}{#2}}}\frontmatter@RRAPformat}
  {}{}
}%
\makeatother
\begin{document}

\preprint{AIP/123-QED}

\title{Terahertz Logic Gate Operations in a Toroidal Metasurface}
% Force line breaks with \\
\author{Angana Bhattacharya}
\affiliation{ Physics Department, Indian Institute of Technology Guwahati, Guwahati, Assam, India, 781039.
}
\email{angana18@iitg.ac.in}
\author{Bhagwat Singh Chouhan}%
 \affiliation{ Physics Department, Indian Institute of Technology Guwahati, Guwahati, Assam, India, 781039.
}
\author{Kajal Sharma}
\affiliation{ Physics Department, Indian Institute of Technology Guwahati, Guwahati, Assam, India, 781039.
}
 \author{Gagan Kumar}
 \affiliation{ Physics Department, Indian Institute of Technology Guwahati, Guwahati, Assam, India, 781039.
}

\date{\today}% It is always \today, today,

\begin{abstract}
The applications of terahertz metamaterials are being actively explored in recent times for applications in high-speed communication devices, miniature photonic circuits, and bio-chemical devices because of their wide advantages. The toroidal resonance, a new type of metasurface resonance, has been examined with great interest to utilize its properties in terahertz metasurface applications. The study reports a proof of concept design of a toroidal metasurface that experimentally demonstrates logic gate operations in the terahertz frequency regime by passive tuning of the split ring resonators compromising the meta-atom design. The amplitude modulation is utilized as a method of determining the Boolean logic output of the system. The proposed metasurface could be further optimized for high amplitude modulations, as well as for active logic gate operations using tunable materials including graphene and ITO. The proposed metasurface consists of three split-ring resonators, and the near-field coupling between the adjacent resonators dictates the logic gate operations. The toroidal excitation in the metasurface is determined by a multipole analysis of the scattered powers of terahertz radiation. The proposed metasurfaces experimentally define AND Boolean logic gate at 0.89 terahertz, and OR Boolean logic gate at 0.97 terahertz. Numerical simulations support the experimentally obtained results. Additionally, we numerically report the excitation of NAND gate at 0.87 THz.  Such toroidal logic-gate metasurfaces in the terahertz region could find applications in digitized terahertz circuits for photonic device applications.
\end{abstract}

\maketitle

\section{Introduction}
Metamaterial (MM) technology has seen large interest, as well as advancements in the last decade because of its numerous applications \cite{1}. MMs are artificially designed structures that consist of arrays of resonant elements, termed as meta-atoms, which mimic the properties of natural materials \cite{2}. The dimensions of the meta-atoms determine the properties of the MM. The meta-atoms are designed smaller than the wavelength of the incident radiation, and hence the non-uniformity of the meta-atom array is lost to the incident radiation and a uniform new material, or artificial material is obtained. The metamaterials designed in 2-D planar surfaces are termed as metasurfaces. The meta-atoms are often in the form of resonant elements termed as split ring resonators. Metallic as well as dielectric metasurfaces have found wide applications in the design of modulators, switches, lenses etc\cite{2,3,4,5,6,62}. A pre-dominant application of metasurface is in the field of terahertz (THz) devices \cite{7}. The terahertz radiation is a segment of the electromagnetic spectrum which is sandwiched between the microwave and infra-red frequencies \cite{8}. Lying at the intersection of optics and photonics technologies, it has remained underutilized due to the lack of natural sources vibrating at THz frequencies\cite{9}. However, MMs, with carefully designed meta-atoms, could exhibit the required terahertz responses and be utilized for various purposes. Researchers have demonstrated numerous applications of THz metasurfaces including electromagnetically induced transparency, terahertz bio-sensors, meta-modulators, filters etc \cite{7}. THz metamaterials have also been utilized to study a certain type of electromagnetic resonance, termed as the toroidal resonance \cite{10}. The toroidal dipole excitation takes place when magnetic moments are aligned in a head-to-tail manner \cite{11}. The toroidal excitation is not evident in natural materials due to the domineering effect of electric and magnetic dipole moments \cite{12}. However, meta-atoms with toroidal symmetries could be designed such that the toroidal dipole excitation demonstrates significant effect in metasurfaces \cite{13,14}. The inherent curiosity to study this new type of resonance has resulted in diverse examination of the properties of toroidal metasurfaces \cite{15,16,17,18,19,20,21,22,23,232}, . Toroidal excitation has demonstrated similar far-field radiation pattern to electric dipole radiation and the interference between the two has demonstrated a non-radiation “anapole” configuration \cite{24}. Further, toroidal resonance has displayed evidence of sharp resonances with high quality factors \cite{19, 25,26,27}. Active tuning of metasurfaces, dual toroidal resonances, toroidal BICs have seen wide interests \cite{19,25}. There is a rising curiosity in the applications and capacities of the toroidal resonance in THz metasurfaces. Considering these aspects, this article studies the logic-gate applications in a toroidal THz metasurface. Logic gates are the basic building blocks of computational circuits. Logic gates make decisions based on a combination of inputs fed into it, with both inputs and outputs in the form of Boolean signals \cite{30}. The Boolean operations are performed using light in optical logic gates \cite{31} . A reconfigurable MEMS Fano resonant metasurface has been demonstrated that performs terahertz logic gate operations with electrical inputs exhibiting XOR and XNOR operations at far field, and NAND operation in the near-field \cite{32}. Further, NOR and OR Boolean operations have been demonstrated in an all optical terahertz logic gate made from a semiconductor-metal hybrid metasurface \cite{33}. THz logic gates have also been demonstrated in metasurfaces based on spoof plasmons, in electro -thermally tunable metasurfaces, and in graphene based metasurfaces \cite{34,35,36,37}. However,toroidal metasurfaces have not been studied to explore their logic gate capabilities to the best of our knowledge. The study of a toroidal logic gate in THz metasurface provides an extra degree of freedom to explore the novel toroidal excitation for computational applications. 
\newline
In this study, we report logic gate operations in a toroidal metasurface by studying its transmission response in the terahertz regime. The carefully designed metasurface consists of three split ring resonators, and the near field coupling between the resonators determines the Boolean state ‘0’(indicating OFF state) and ‘1’ ( indicating ON state) in the metasurface. A thorough understanding of the logic gate operations is provided by experimental results and supported by numerical simulations. The proposed geometry demonstrates low amplitude modulation among the Boolean states, but the design can be further optimized to  show higher amplitude modulations for the different Boolean states by incorporating asymmetries and exciting fano or BIC type resonances. The proposed design serves as an experimental proof of concept system for extending the idea of toroidal logic gates into active metasurfaces and passive modulations with higher amplitude depth. Further, the idea can be extended to other frequency domains by tweaking the toroidal metasurface geometry. The article is arranged as follows. We initially discuss the design of the metasurface, its experimental fabrication and THz characterisation and the excitation of the toroidal resonance in the metasurface. This is followed by the analysis and discussion of logic gate operations in the metasurface experimentally and numerically . The conclusion section provides a summary of the experimental and numerical findings of the study, the associated drawbacks, scope of improvements, and future prospects of the study .

\section{Metamaterial Design and Simulation}

\begin{figure}
	\includegraphics[scale=0.25]{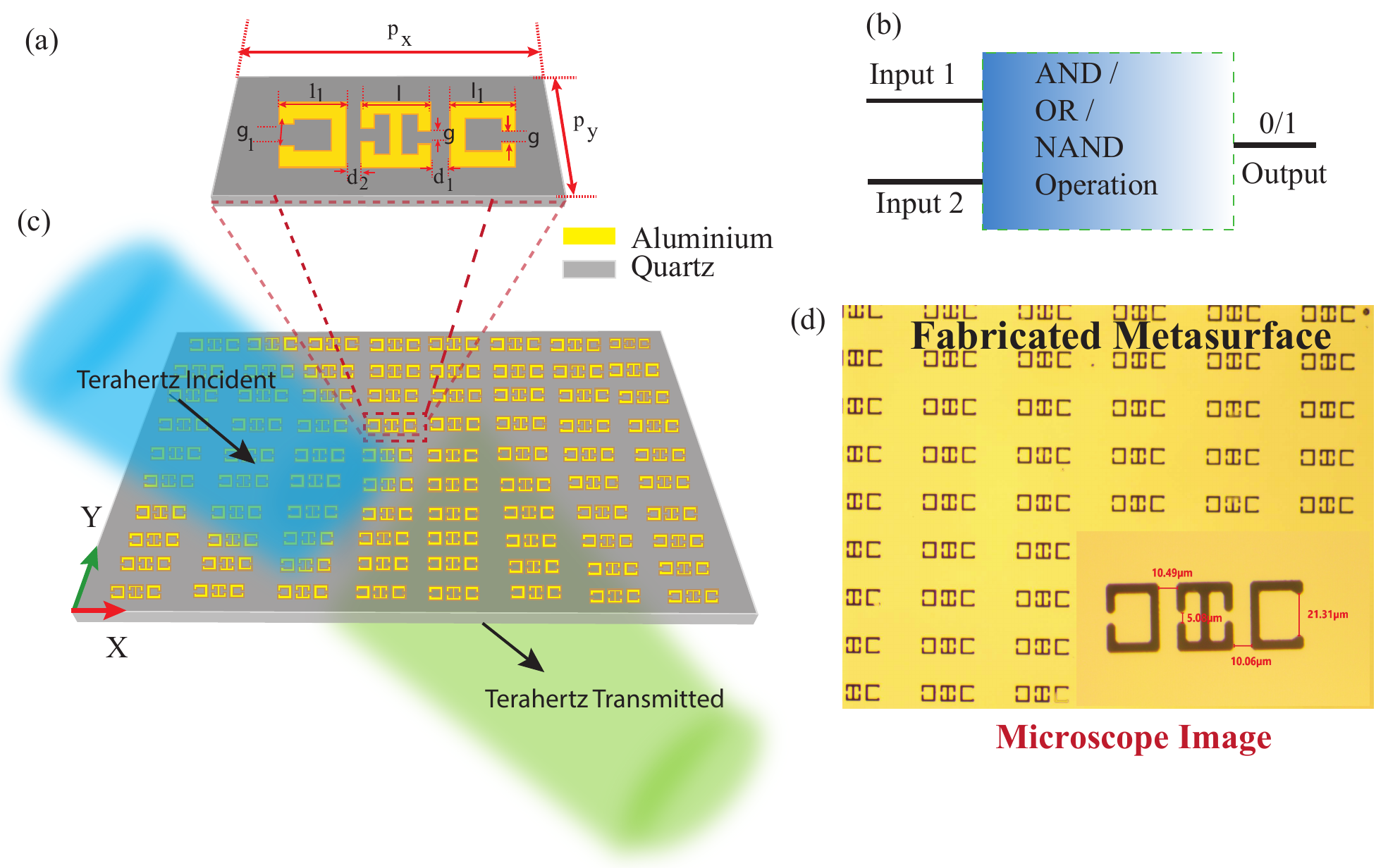}\centering
	\caption{
			a) Schematic of the logic gate meta-atom having periodicity along x and y direction as $p_{x}$ and $p_{y}$ respectively, length of resonator ‘l’, capacitive gap ‘g’, and distance between adjacent resonators $d_{1}$, and $d_{2}$. b) A schematic for the two- input logic gate operation in the MM. c) A schematic of the metasurface array. Y-polarized THz radiation is incident normally on the metasurface. (d) Microscope image of the fabricated metasurface array. A single meta-atom is shown in a magnified view.
		 }
\end{figure}
The basic design of the metasurface was inspired by a previously studied geometry and consists of three aluminium meta-atoms in a quartz substrate[22]. However, adequate modifications have been made keeping in mind the applications of the metasurface. Figure 1 (a) shows the schematic of a meta-atom of the proposed metasurface. The design was chosen for its simplicity and ease of fabrication. The central split ring resonator (SRR) demonstrates toroidal nature while the left and right C resonator shows a dipolar nature. The periodicity along the x direction is $p_x$= 200 $\mu$m, while the periodicity along the y direction is $p_y$= 100 $\mu$m. The length of the central resonator was chosen as l=35 $\mu$m and that of the C resonators was $l_1$=32 $\mu$m. The width of the resonators was 5 µm. The capacitive gaps on the central resonators g is 5 $\mu$m. The capacitive gap on the right resonator is, g= 5 $\mu$m. The capacitive gap on the left resonator is, $g_1$= 21$\mu$m. The breadth of all the SRRs is, b = 35 $\mu$m. The distance between the central resonator and C-shaped resonator on the right-hand side is termed as \textquoteleft $d_1$\textquoteright and that between the central and left \textquoteleft C \textquoteright resonator is termed as \textquoteleft $d_{2}$\textquoteright . The variation of $d_1$  and $d_2$ is analysed to study the logic gate operation in our MM sample. The design of the metasurface and numerical simulations were performed using CST Microwave Studio simulation software. Figure 1(c) shows the schematic of the metasurface array on which terahertz (THz) radiation is incident normally, with polarisation parallel to the capacitive gaps. Figure 1 (b) demonstrates a block diagram of the logic gate operations that could be undertaken via the MM. A 2-input logic gate provides a Boolean output on the basis of the combination of signals fed in the input. Figure 1 (d) shows a microscope image of the experimentally fabricated metasurface using photolithography in a clean room environment. The image is in a mirror image format as observed in the microscope. The image of a single meta-atom is shown in magnified view for better clarity and understanding.
\section{Fabrication and Characterization of the Terahertz Metasurface}
The metasurface was fabricated using traditional photolithography technique. A quartz substrate ($\varepsilon$=3.7) was initially cleaned using acetone and further, coated with a 200 nm thick aluminium layer using thermal deposition technique. Further, it was spincoated with positive photoresist (S1813) and the desired pattern was transferred to the quartz substrate using UV radiation and a hard mask. The sample was then developed and the 1cm-by-1cm array of meta-atoms were obtained. The final metasurface configuration was fabricated for four different geometries for values of of $d_1$  and $d_2$ corresponding to (10 $\mu$m, 10 $\mu$m), (10 $\mu$m, 20 $\mu$m), (20 $\mu$m, 10 $\mu$m), and (20 $\mu$m, 20 $\mu$m) respectively. The fabricated samples were then characterized using a THz-time domain spectroscopy setup consisting of two fibre-coupled photo-conductive antennas\cite{38}. The response of the bare quartz substrate, termed as reference, is first analysed followed by the response of the metasurface, termed as sample. The ratio of the sample response to the reference response provides the transmission through the metasurface.

\section{Results and Discussion}
\begin{figure}
	\includegraphics[scale=0.35]{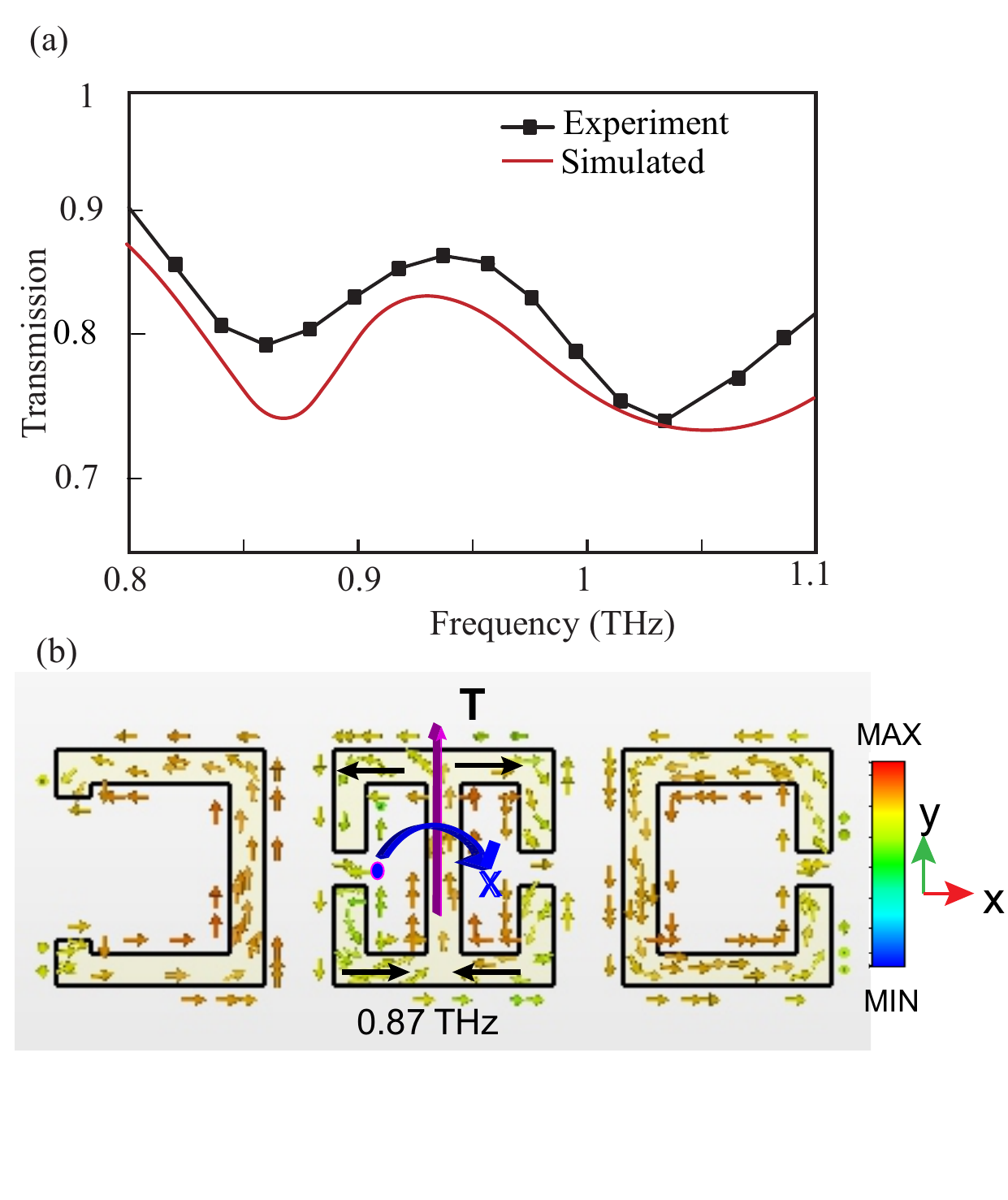}\centering
\caption{(a) Simulated and experimentally measured terahertz transmission response of the metasurface for the $d_1$ = $d_2$ = 10 $\mu$m configuration. (b) Surface current profile of the metasurface for the resonance at 0.87 THz. Toroidal dipole excitation is shown along the Y-axis.}
\end{figure}
The THz response of the metasurface is studied via simulations, as well as experimental measurements of the transmission spectrum. The response is measured for a configuration where the distance between the adjacent resonators is $d_1$ = $d_2$ = 10 $\mu$m. The transmission spectrum is shown in figure 2. The red line in figure 2 (a) shows the simulated transmission spectrum using CST Microwave Studio software, and black curve in figure 2 (a) illustrates the experimentally measured transmission spectrum measured using the THZ-TDS setup. From the simulated transmission spectrum, we observe two resonance dips at 0.87 THz and 1.07 THz. The THz transmission through the fabricated metasurface is measured and demonstrates two resonances and a close match between the simulated and experimentally measured results is observed. In this study, we have focused on the first resonance at 0.87 THz. The surface current profile of the metasurface at 0.87 THz is analyzed, as shown in figure 2 (b). It may be observed that, for the central resonator, the surface current profile flows clockwise on the right-hand arm and flows anticlockwise on the left-hand arm of the resonator. The clockwise flow of current on the right arm leads to a magnetic moment going inside the plane on the right side (indicated by the cross symbol), while the anticlockwise flow on the left arm leads to magnetic moment coming outside the plane on the left side of the resonator (indicated by the dot symbol). This end-to-end formation of magnetic moments lead to the excitation of a toroidal dipole moment along the y direction. Thus, there is a clear excitation of toroidal dipole moment in the metasurface at the resonant frequency of 0.87 THz. 
\newline
\begin{figure}
	\includegraphics[scale=0.3]{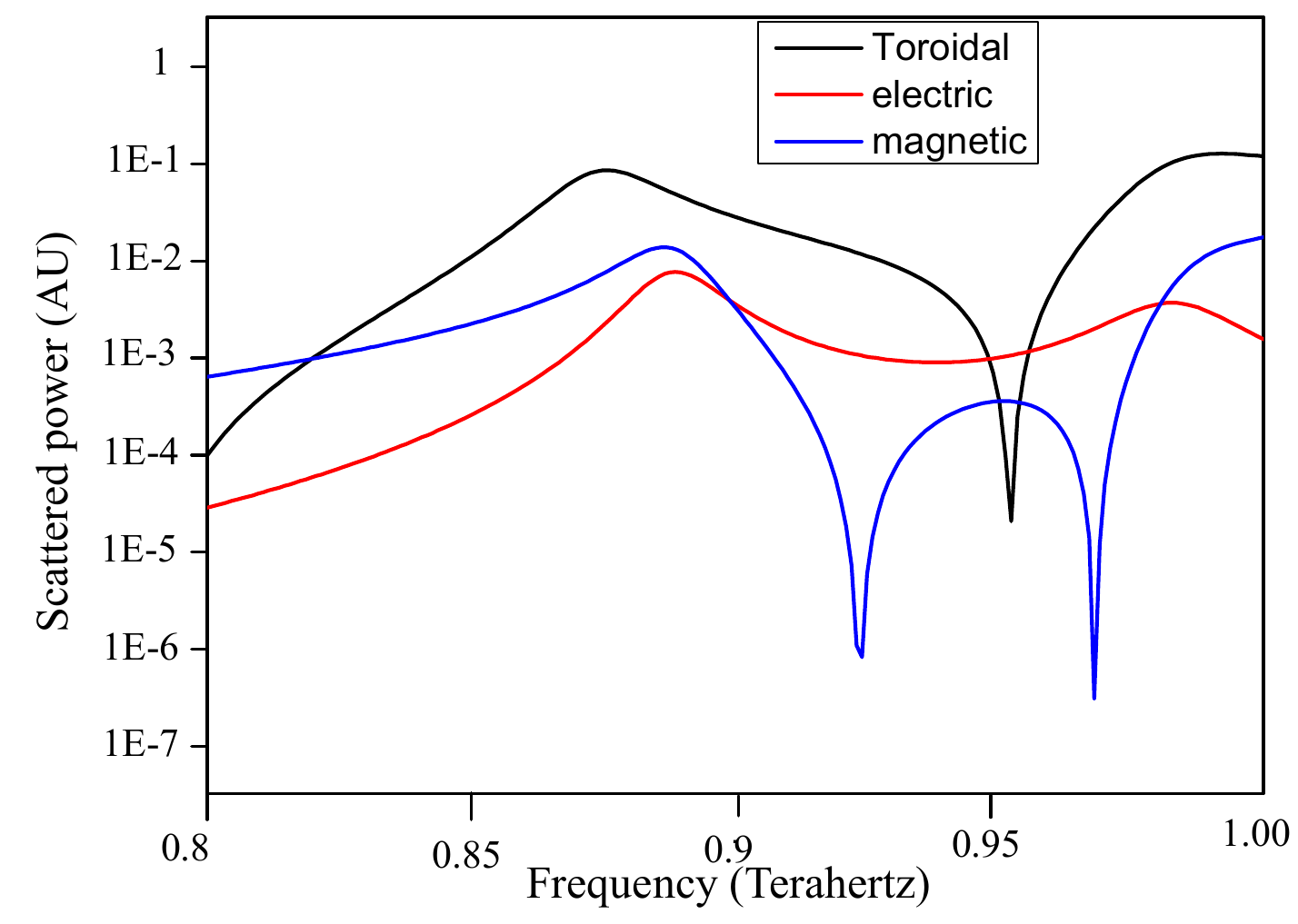}\centering
	\caption{Multipole analysis of scattered power by electric, magnetic, and toroidal dipole moments over the THz frequency range.}
\end{figure}
Further, we study the nature of electromagnetic resonances in the metamaterial structure. A multipolar analysis is performed over the terahertz frequency range of 0.8 THz to 1 THz and the power scattered by electric, magnetic, and toroidal dipolar excitation is plotted for the $d_1$= 10 $\mu$m, $d_2$=10 $\mu$m configuration, as shown in figure 3. The multipole analysis has been studied in several articles wherein the power scattered by the different electromagnetic moments have been evaluated theoretically. The black line in figure 3 indicates toroidal scattered power, the blue line indicates power scattered by magnetic dipole moment, and the red line indicates electric dipole scattered power. It is seen that at 0.87 THz where the metasurface has its first resonance, there is highest toroidal scattered power. This confirms the toroidal behaviour of the metasurface. It is also observed that in the frequency range of 0.87 THz to 0.95 THz, and also in the range of 0.96 to 1 THz, there is toroidal highest toroidal dipole scattered power. Thus, the window containing the frequencies that demonstrate logic gate behaviour has pre-dominantly toroidal scattered power.
\section{Study of AND and OR Logic Gates}
The study of logic gate operation is done via the change in the distance between the central and the other ‘C’ shaped resonators.  The initial configuration that was analyzed had a symmetrical structure with both the C resonators at a distance $d_1$=$d_2$=10 $\mu$m from the central SRR. The transmission spectrum for this symmetric configuration is discussed in the previous section, and shown in figure 2. We consider this configuration as the Boolean input \textquoteleft11\textquoteright configuration. The displacement of the left-side side resonator from the central resonator by a unit of $d_2$=20 $\mu$m, while the right-hand side resonator is fixed at $d_1$=10 $\mu$m, is considered as a Boolean input \textquoteleft 10\textquoteright. This variation in the transmission for varying the distances of the C shaped resonators from the central resonator is shown in figure 4. The experimentally measured transmission spectrums for the fabricated metasurfaces of varying $d_1$,and $d_2$ is depicted in figure 4(a). The transmission spectrum for the $\textquoteleft10\textquoteright$ configuration is shown in figure 4(a) by the red curve.  A blue shift in the first resonance dip is observed as it is shifted to 0.9 THz as compared to the 11 configuration. The configuration where the right SRR is at a distance $d_1$=20 $\mu$m from the central resonator, while $d_2$ is fixed at 10 $\mu$m is termed as the \textquoteleft01\textquoteright input condition. For this configuration the resonance dips are observed at 0.89 THz and at 1.05 THz as can be seen via the blue line in figure 4(a). Further, the configuration where both the C resonators are separated from the central resonator by distance $d_1$=$d_2$=20 $\mu$m is termed as the \textquoteleft00\textquoteright configuration. We term this as the \textquoteleft00\textquoteright input as it is believed that increasing the distance between the central resonator and the C resonators will lead to a reduced near field coupling between the SRRs. Hence, the decrease in the near field coupling is assumed to correspond to the \textquoteleft OFF\textquoteright state or 00 state in our study. Similarly, we believe that the $d_1$=$d_2$=10 $\mu$m input state is the \textquoteleft ON\textquoteright state, or \textquoteleft 11\textquoteright state as both the C-SRRs will strongly couple to the middle resonator. For the 00 configuration we observe one resonance dip at approximately 0.89 THz and a broad resonance dip at 1.05 THz as shown by the pink curve of figure 4(a). On varying the distances $d_1$ and $d_2$, a variation in a transmission amplitude and resonance width is observed. The resonances becomes broader and there is an increase in the transmission depth as the distances are varied.  It is evident that all the resonance dips and peaks of the transmission spectrum occur within a small window of 0.8 THz to 1.1 THz. For our study of logic gate operation in a MM, we intend to utilize this narrow THz window to explore the various logic gates in the metasurface.
\begin{figure}
	\includegraphics[scale=0.45]{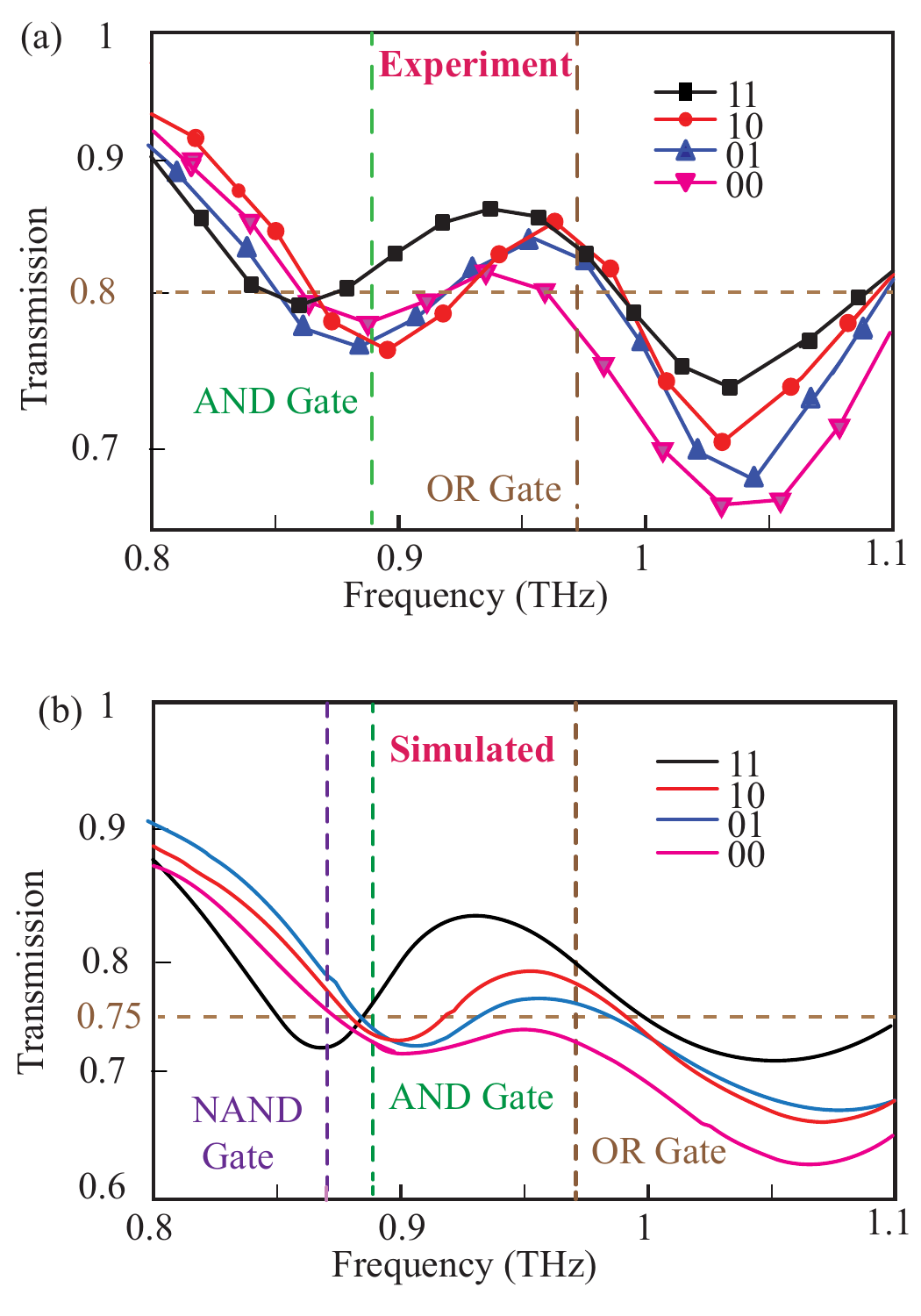}\centering
	\caption{(a) Proof of concept for logic gate operations for various input states of the metasurface at varying frequencies. A transmission amplitude of 80 \% and higher is taken as the cutoff for logic output state 1 (ON state) for experimental measurements, while a cutoff of 75 \% is taken for simulated results.  The input states: 00, 01, 10, 11 are determined by the passive displacement of the C-resonators from the central resonator. a) Logic gate operation at 0.89 THz (AND) and 0.97 THz (OR)for experimentally measured results. (b) Logic gate operation at 0.87 THz (NAND), 0.89 THz (AND), and 0.97 THz (OR) for numerically simulated results. }
\end{figure}
In order to realise the logic gate operations, we scrutinize the different regions that could exhibit logic gate output for the specified inputs for the window of 0.8 THz to 1.1 THz, as shown in figure 2. In this study, we fix a transmission amplitude of 80 \% to qualify for a Boolean output 1 state. A transmission amplitude less than 80 \% is assumed to be a Boolean output 0 state. The black curve in figure 4 indicates the input 11 state, the red curve indicates input 10 state, the blue curve indicates 01 state, and the pink curve indicates 00 state, as discussed. It is observed that at 0.89 THz, the 11 state has a transmission higher than 80 \% while the 10, 01, and 00 states demonstrate a transmission amplitude of less than 80 \%. The truth table corresponding to this configuration at 0.89 THz is described in table 1 (a). The truth table output for all states besides the 11 state is Boolean 0, while the output for the 11 state is Boolean 1. From the truth table 1 (a), it can be understood that the metasurface at 0.89 THz corresponds to a Boolean AND operation. Further, the response of the metasurface at 0.97 THz is analysed. The black, red, and blue curves corresponding to the 11, 10, and 01 input states respectively have transmission higher than 80 \%, while the pink curve corresponding to 00 input state has a transmission less than 80 \% at 0.97 THz as can be seen in figure 4. The corresponding truth table is depicted in table 1(b). The output of the truth table at 0.97 THz corresponds to the OR gate configuration. Thus, the metasurface experimentally demonstrates AND, and OR logic gate operations at 0.89 THz and 0.97 THz respectively. 
\newline
The experimental results are further verified using numerical simulations by studing the transmission spectrum of the metasurface for the different configurations of $d_1$ and $d_2$. The simulation results are depicted in figure 4(b). The black curve in figure 4(b) illustrates the (1,1) configuration. The first resonance for the black curve lies at 0.87 THz and the second resonance dip is at approximately 1.05 THz. Further, the transmission spectrum corresponding to the other variations of $d_1$ and $d_2$ are also analysed. The red curve in figure 4(b) corresponds to the simulated (1,0) configuration. The blue curve corresponds to the simulated (0,1) configuration, while the pink curve in figure 4(b) corresponds to the simulated (0,0) configuration. It may be observed that there is some small mismatch between the amplitude of the simulated transmission dips as compared to that of the experimentally measured curves depicted in figure 4(a). These differences in the amplitude of the transmission dips may be attributed to the fabrication imperfections of the fabricated metasurfaces, environmental effects, diffraction losses, and due to the difference in the resolution of the simulation and measured results using the THz-TDS setup. To take into account the shift in the transmission amplitude for the simulation results, a cutoff of 75 \% transmission is set for the Boolean output 1 in the simulated results. From figure 4(b), it can be observed that at 0.89 THz, the black curve corresponding to Boolean input (1,1) is higher than the cutoff transmission, leading to a Boolean output of 1. The other curves corresponding to input configurations (1,0), (0,1), and (0,0) are below the cut-off transmission of 75 \%, hence leading to a Boolean output of 0. The truth table for the simulated curve at 0.89 THz matches with the truth table 1 (a) for AND gate, illustrated for the experimentally measured results at 0.89 THz. Thus, at 0.89 THz by simulation, AND gate is obtained. Next, the response at 0.97 THz is analyzed and it is observed that the simulated curves have transmission higher than the 75 \% cutoff for for all input configurations, corresponding to a Boolean output 1, except the (0,0) configuration which corresponds to a Boolean output 0. This is matched with the truth table for OR gate as depicted in truth table 1(b). Thus, at 0.97 THz, simulation results match with experimental results and OR gate is obtained. Hence, the metasurface demonstrates AND and OR logic gate.
\\
\\
\textbf{NAND Gate Operation }
\\
\\
Further, for the simulated transmission spectrum, the response at 0.87 THz was analyzed and it was observed that transmission is higher than the cutoff value of 75 \% for the (01), (10), and (0,0) configurations, while it is lower than the cutoff for the (1,1) configuration. The truth table corresponding to this observation is shown in table 1(c). The response at 0.87 THz corresponds to a Boolean output of NAND gate configuration. Thus, via simulation, the metasurface demonstrates OR, AND, and NAND gates at 0.97 THz, 0.89 THz, and 0.87 THz respectively.
The truth tables for the experimentally analysed logic gates at 0.89 THz, 0.97 THz, and 0.87 THz are illustrated below. The two inputs, input 1 and input 2, depict the distance $d_1$, and $d_2$, with the input being \textquotedblleft high\textquotedblright or \textquoteleft1 \textquoteright for  $d_1$ or $d_2$ = 10 $\mu$m, and the input being \textquotedblleft low\textquotedblright or \textquoteleft 0\textquoteright for $d_1$ or $d_2$ = 20 $\mu$m.

\begin{table}
\caption{Truth table for different combination of input states at varying frequencies ‘f’. a) Truth table for AND Gate at 0.89 THz. b) Truth table for OR Gate at 0.97 THz. c) Truth table for NAND Gate at 0.87 THz.}
 %\subfloat{\textbf{a) f = 0.89 THz (AND Gate)}}
	%\subcaptionbox{f = 0.89 THz (AND Gate)}{
 \begin{subtable}{.5\linewidth}\centering
		\begin{tabular}[t]{|c|c|c|}
            \hline
			Input 1 ($d_{1}$)&Input 2 ($d_{2}$)&Output\\ \hline
			0&0&0 \\ \hline
			0&1&0\\ \hline
			1&0&0 \\ \hline   
			1&1&1\\ \hline 
		\end{tabular}
	%}
\caption{f = 0.89 THz (AND Gate)}\label{tab:1a}
\end{subtable}
%\subfloat{\textbf{a) f = 0.89 THz (AND Gate)}}
% \begin{subtable}	
 %\subcaptionbox{f=0.97 THz (OR Gate)}{
 \begin{subtable}{.5\linewidth}\centering
		\begin{tabular}[t]{|c|c|c|}
			\hline
			Input 1 ($d_{1}$)&Input 2 ($d_{2}$)&Output\\ \hline
			0&0&0 \\ \hline
			0&1&1\\ \hline
			1&0&1 \\ \hline   
			1&1&1\\ \hline 
		\end{tabular}
\caption{f=0.97 THz (OR Gate)}\label{tab:1a}
\end{subtable}
%}
% \vspace{3mm}
 % \centering
  %\end{subtable}
%\begin{subtable}	
 %\subcaption{\textbf{c) f = 0.89 THz (AND Gate)}}
	%\subcaptionbox{f= 0.87 THz (NAND Gate)}{
  \begin{subtable}{.5\linewidth}\centering
		\begin{tabular}[t]{|c|c|c|}
			\hline
			Input 1 ($d_{1}$)&Input 2 ($d_{2}$)&Output\\ \hline
			0&0&1 \\ \hline
			0&1&1\\ \hline
			1&0&1 \\ \hline   
			1&1&0\\ \hline 
		\end{tabular}
	%}
\caption{f= 0.87 THz (NAND Gate)}\label{tab:1b}
\end{subtable}	
\end{table}

The near field coupling between the resonators was also studied for the different metasurface configurations at the first resonance.Figure 5 demonstrates the near field coupling between the resonators for varying distances $d_1$ and $d_2$. Figure 5(a) shows the $d_1$=$d_2$=10 $\mu$m configuration, and it may be observed that all three resonators are excited and there is strong near-field coupling between the SRRs which verify our consideration of the $d_1$=$d_2$=10 $\mu$m configuration as the “11” Boolean input. We next studied the electric field coupling for the $d_1$=$d_2$=20 $\mu$m configuration. We observe that the three resonators are individually excited by the y-polarized THz radiation, however there is reduced near field coupling between the resonators, as can be seen from figure 5(b), and we believe this reduced coupling validates our consideration of the $d_1$=$d_2$=20 $\mu$m configuration as the “00” Boolean input. The interaction between the SRRs for $d_1$= 10 $\mu$m,  $d_2$=20 $\mu$m configuration is shown in figure 5(c).
\begin{figure}
	\includegraphics[scale=0.27]{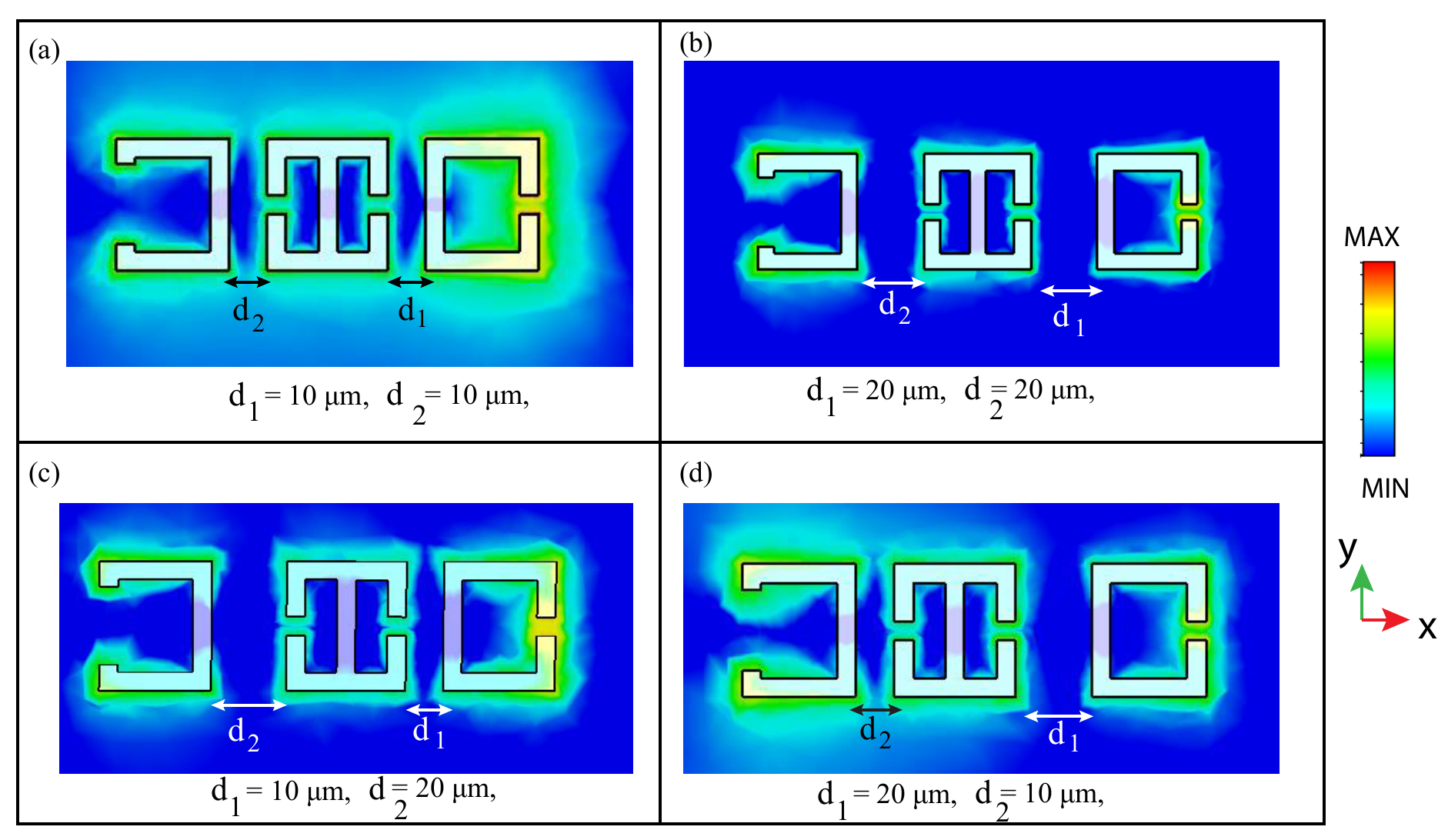}\centering
	\caption{The electric field excited in the four metasurface configurations and analysis of their near field coupling. (a) Electric field profile for the $d_1$=$d_2$=10 $\mu$m metasurface configuration. (b) Electric field profile for the $d_1$=$d_2$=20 $\mu$m metasurface configuration. (c) Electric field profile for the $d_1$=10 $\mu$m $d_2$=20 $\mu$m metasurface configuration. (d) Electric field profile for the $d_1$=20 $\mu$m, $d_2$=10 $\mu$m metasurface configuration.}
\end{figure}
It may be observed that there is stronger near field confinement between the central and right resonators, and weak interaction between the central and left resonators and thus, we consider the $d_1$= 10 $\mu$m, $d_2$=20 $\mu$m configuration as the “10” Boolean input. The near field interaction in the $d_1$= 20 $\mu$m, $d_2$=10 $\mu$m configuration is shown in figure 5(d), and similar to the case of figure 5(c), stronger near field interaction is observed between the central and left resonator as compared to the central and right resonator. Thus, this verifies our consideration of the $d_1$= 20 $\mu$m, $d_2$=10 $\mu$m configuration as the Boolean “01” input.

\section{Conclusions}
The study explores the design of a toroidal metasurface demonstrating logic gate operations in the range of 0.85 THz to 1 THz. It is confirmed via electric field profiles that there is strong bright-bright mode coupling between the SRRs when they are placed close together at a distance of $d_1$=$d_2$=10 $\mu$m, which is assumed to be the \textquoteleft11\textquoteright Boolean input state. Further, the reduced coupled state corresponding to $d_1$=$d_2$=20 $\mu$m is taken as the \textquoteleft00\textquoteright  input state. The variation of distances $d_1$ and $d_2$ leads to shifts in the transmission spectrum and the displacements are analyzed to demonstrate logic gate operations at different frequencies. We report the excitation of AND and OR logic gates experimentally and numerically in the metasurface at 0.89 THz and 0.97 THz respectively. Further, numerical simulations demonstrate NAND logic gate at 0.87 THz. The numerical and experimental measurements show a close match in the results. The near field coupling between the resonators is also analyzed to explain the logic gate input configurations. Further, a multipole analysis is performed over the frequency range of 0.8 THz to 1 THz and toroidal behaviour at the resonance is confirmed for the metasurface at the logic-gate frequencies. The study serves as a proof of concept idea to demonstrate and further improve the functioning of toroidal logic gate devices in the terahertz domain. There is further scope in the study via active modulation of parameters to demonstrate an active toroidal logic gate in the terahertz region.  The disadvantage of low amplitude modulation for the Boolean ON and OFF states could be overcome by optimized metasurface designs having sharper toroidal resonance with increased transmission depth. Further, the passive nature of the logic gate operations could be improved by introducing a graphene layer to demonstrate active toroidal logic gate operations. Such toroidal logic gate metasurfaces could be significant in the design of futuristic terahertz compact devices and on-chip photonic circuits.

\begin{acknowledgments}
Author GK would like to acknowledge the financial support from SERB (CRG/2019/002807).
\end{acknowledgments}

\section*{Data Availability Statement}
The data that support the findings of this study are available from the corresponding author upon reasonable request.
\section{References}
\nocite{*}
\bibliography{ref}% Produces the bibliography via BibTeX.

\begin{thebibliography}{40}%
\makeatletter
\providecommand \@ifxundefined [1]{%
 \@ifx{#1\undefined}
}%
\providecommand \@ifnum [1]{%
 \ifnum #1\expandafter \@firstoftwo
 \else \expandafter \@secondoftwo
 \fi
}%
\providecommand \@ifx [1]{%
 \ifx #1\expandafter \@firstoftwo
 \else \expandafter \@secondoftwo
 \fi
}%
\providecommand \natexlab [1]{#1}%
\providecommand \enquote  [1]{``#1''}%
\providecommand \bibnamefont  [1]{#1}%
\providecommand \bibfnamefont [1]{#1}%
\providecommand \citenamefont [1]{#1}%
\providecommand \href@noop [0]{\@secondoftwo}%
\providecommand \href [0]{\begingroup \@sanitize@url \@href}%
\providecommand \@href[1]{\@@startlink{#1}\@@href}%
\providecommand \@@href[1]{\endgroup#1\@@endlink}%
\providecommand \@sanitize@url [0]{\catcode `\\12\catcode `\$12\catcode
  `\&12\catcode `\#12\catcode `\^12\catcode `\_12\catcode `\%12\relax}%
\providecommand \@@startlink[1]{}%
\providecommand \@@endlink[0]{}%
\providecommand \url  [0]{\begingroup\@sanitize@url \@url }%
\providecommand \@url [1]{\endgroup\@href {#1}{\urlprefix }}%
\providecommand \urlprefix  [0]{URL }%
\providecommand \Eprint [0]{\href }%
\providecommand \doibase [0]{http://dx.doi.org/}%
\providecommand \selectlanguage [0]{\@gobble}%
\providecommand \bibinfo  [0]{\@secondoftwo}%
\providecommand \bibfield  [0]{\@secondoftwo}%
\providecommand \translation [1]{[#1]}%
\providecommand \BibitemOpen [0]{}%
\providecommand \bibitemStop [0]{}%
\providecommand \bibitemNoStop [0]{.\EOS\space}%
\providecommand \EOS [0]{\spacefactor3000\relax}%
\providecommand \BibitemShut  [1]{\csname bibitem#1\endcsname}%
\let\auto@bib@innerbib\@empty
%</preamble>
\bibitem [{\citenamefont {Zheludev}\ and\ \citenamefont {Kivshar}(2012)}]{1}%
  \BibitemOpen
  \bibfield  {author} {\bibinfo {author} {\bibfnamefont {N.~I.}\ \bibnamefont
  {Zheludev}}\ and\ \bibinfo {author} {\bibfnamefont {Y.~S.}\ \bibnamefont
  {Kivshar}},\ }\bibfield  {title} {\enquote {\bibinfo {title} {From
  metamaterials to metadevices},}\ }\href@noop {} {\bibfield  {journal}
  {\bibinfo  {journal} {Nature materials}\ }\textbf {\bibinfo {volume} {11}},\
  \bibinfo {pages} {917--924} (\bibinfo {year} {2012})}\BibitemShut {NoStop}%
\bibitem [{\citenamefont {Liu}\ and\ \citenamefont {Zhang}(2011)}]{2}%
  \BibitemOpen
  \bibfield  {author} {\bibinfo {author} {\bibfnamefont {Y.}~\bibnamefont
  {Liu}}\ and\ \bibinfo {author} {\bibfnamefont {X.}~\bibnamefont {Zhang}},\
  }\bibfield  {title} {\enquote {\bibinfo {title} {Metamaterials: a new
  frontier of science and technology},}\ }\href@noop {} {\bibfield  {journal}
  {\bibinfo  {journal} {Chemical Society Reviews}\ }\textbf {\bibinfo {volume}
  {40}},\ \bibinfo {pages} {2494--2507} (\bibinfo {year} {2011})}\BibitemShut
  {NoStop}%
\bibitem [{\citenamefont {Zouhdi}, \citenamefont {Sihvola},\ and\ \citenamefont
  {Vinogradov}(2008)}]{3}%
  \BibitemOpen
  \bibfield  {author} {\bibinfo {author} {\bibfnamefont {S.}~\bibnamefont
  {Zouhdi}}, \bibinfo {author} {\bibfnamefont {A.}~\bibnamefont {Sihvola}}, \
  and\ \bibinfo {author} {\bibfnamefont {A.~P.}\ \bibnamefont {Vinogradov}},\
  }\href@noop {} {\emph {\bibinfo {title} {Metamaterials and plasmonics:
  fundamentals, modelling, applications}}}\ (\bibinfo  {publisher} {Springer
  Science \& Business Media},\ \bibinfo {year} {2008})\BibitemShut {NoStop}%
\bibitem [{\citenamefont {Caloz}(2009)}]{4}%
  \BibitemOpen
  \bibfield  {author} {\bibinfo {author} {\bibfnamefont {C.}~\bibnamefont
  {Caloz}},\ }\bibfield  {title} {\enquote {\bibinfo {title} {Perspectives on
  em metamaterials},}\ }\href@noop {} {\bibfield  {journal} {\bibinfo
  {journal} {Materials Today}\ }\textbf {\bibinfo {volume} {12}},\ \bibinfo
  {pages} {12--20} (\bibinfo {year} {2009})}\BibitemShut {NoStop}%
\bibitem [{\citenamefont {Xiao}\ \emph {et~al.}(2020)\citenamefont {Xiao},
  \citenamefont {Wang}, \citenamefont {Liu}, \citenamefont {Zhou},
  \citenamefont {Jiang},\ and\ \citenamefont {Zhang}}]{5}%
  \BibitemOpen
  \bibfield  {author} {\bibinfo {author} {\bibfnamefont {S.}~\bibnamefont
  {Xiao}}, \bibinfo {author} {\bibfnamefont {T.}~\bibnamefont {Wang}}, \bibinfo
  {author} {\bibfnamefont {T.}~\bibnamefont {Liu}}, \bibinfo {author}
  {\bibfnamefont {C.}~\bibnamefont {Zhou}}, \bibinfo {author} {\bibfnamefont
  {X.}~\bibnamefont {Jiang}}, \ and\ \bibinfo {author} {\bibfnamefont
  {J.}~\bibnamefont {Zhang}},\ }\bibfield  {title} {\enquote {\bibinfo {title}
  {Active metamaterials and metadevices: a review},}\ }\href@noop {} {\bibfield
   {journal} {\bibinfo  {journal} {Journal of Physics D: Applied Physics}\
  }\textbf {\bibinfo {volume} {53}},\ \bibinfo {pages} {503002} (\bibinfo
  {year} {2020})}\BibitemShut {NoStop}%
\bibitem [{\citenamefont {Kildishev}, \citenamefont {Boltasseva},\ and\
  \citenamefont {Shalaev}(2013)}]{6}%
  \BibitemOpen
  \bibfield  {author} {\bibinfo {author} {\bibfnamefont {A.~V.}\ \bibnamefont
  {Kildishev}}, \bibinfo {author} {\bibfnamefont {A.}~\bibnamefont
  {Boltasseva}}, \ and\ \bibinfo {author} {\bibfnamefont {V.~M.}\ \bibnamefont
  {Shalaev}},\ }\bibfield  {title} {\enquote {\bibinfo {title} {Planar
  photonics with metasurfaces},}\ }\href@noop {} {\bibfield  {journal}
  {\bibinfo  {journal} {Science}\ }\textbf {\bibinfo {volume} {339}},\ \bibinfo
  {pages} {1232009} (\bibinfo {year} {2013})}\BibitemShut {NoStop}%
\bibitem [{\citenamefont {Quevedo-Teruel}\ \emph {et~al.}(2019)\citenamefont
  {Quevedo-Teruel}, \citenamefont {Chen}, \citenamefont {D{\'\i}az-Rubio},
  \citenamefont {Gok}, \citenamefont {Grbic}, \citenamefont {Minatti},
  \citenamefont {Martini}, \citenamefont {Maci}, \citenamefont {Eleftheriades},
  \citenamefont {Chen} \emph {et~al.}}]{62}%
  \BibitemOpen
  \bibfield  {author} {\bibinfo {author} {\bibfnamefont {O.}~\bibnamefont
  {Quevedo-Teruel}}, \bibinfo {author} {\bibfnamefont {H.}~\bibnamefont
  {Chen}}, \bibinfo {author} {\bibfnamefont {A.}~\bibnamefont
  {D{\'\i}az-Rubio}}, \bibinfo {author} {\bibfnamefont {G.}~\bibnamefont
  {Gok}}, \bibinfo {author} {\bibfnamefont {A.}~\bibnamefont {Grbic}}, \bibinfo
  {author} {\bibfnamefont {G.}~\bibnamefont {Minatti}}, \bibinfo {author}
  {\bibfnamefont {E.}~\bibnamefont {Martini}}, \bibinfo {author} {\bibfnamefont
  {S.}~\bibnamefont {Maci}}, \bibinfo {author} {\bibfnamefont {G.~V.}\
  \bibnamefont {Eleftheriades}}, \bibinfo {author} {\bibfnamefont
  {M.}~\bibnamefont {Chen}},  \emph {et~al.},\ }\bibfield  {title} {\enquote
  {\bibinfo {title} {Roadmap on metasurfaces},}\ }\href@noop {} {\bibfield
  {journal} {\bibinfo  {journal} {Journal of Optics}\ }\textbf {\bibinfo
  {volume} {21}},\ \bibinfo {pages} {073002} (\bibinfo {year}
  {2019})}\BibitemShut {NoStop}%
\bibitem [{\citenamefont {He}\ \emph {et~al.}(2021)\citenamefont {He},
  \citenamefont {He}, \citenamefont {Dong}, \citenamefont {Wang}, \citenamefont
  {Fu},\ and\ \citenamefont {Zhang}}]{7}%
  \BibitemOpen
  \bibfield  {author} {\bibinfo {author} {\bibfnamefont {J.}~\bibnamefont
  {He}}, \bibinfo {author} {\bibfnamefont {X.}~\bibnamefont {He}}, \bibinfo
  {author} {\bibfnamefont {T.}~\bibnamefont {Dong}}, \bibinfo {author}
  {\bibfnamefont {S.}~\bibnamefont {Wang}}, \bibinfo {author} {\bibfnamefont
  {M.}~\bibnamefont {Fu}}, \ and\ \bibinfo {author} {\bibfnamefont
  {Y.}~\bibnamefont {Zhang}},\ }\bibfield  {title} {\enquote {\bibinfo {title}
  {Recent progress and applications of terahertz metamaterials},}\ }\href@noop
  {} {\bibfield  {journal} {\bibinfo  {journal} {Journal of Physics D: Applied
  Physics}\ }\textbf {\bibinfo {volume} {55}},\ \bibinfo {pages} {123002}
  (\bibinfo {year} {2021})}\BibitemShut {NoStop}%
\bibitem [{\citenamefont {Kleine-Ostmann}\ and\ \citenamefont
  {Nagatsuma}(2011)}]{8}%
  \BibitemOpen
  \bibfield  {author} {\bibinfo {author} {\bibfnamefont {T.}~\bibnamefont
  {Kleine-Ostmann}}\ and\ \bibinfo {author} {\bibfnamefont {T.}~\bibnamefont
  {Nagatsuma}},\ }\bibfield  {title} {\enquote {\bibinfo {title} {A review on
  terahertz communications research},}\ }\href@noop {} {\bibfield  {journal}
  {\bibinfo  {journal} {Journal of Infrared, Millimeter, and Terahertz Waves}\
  }\textbf {\bibinfo {volume} {32}},\ \bibinfo {pages} {143--171} (\bibinfo
  {year} {2011})}\BibitemShut {NoStop}%
\bibitem [{\citenamefont {Koch}\ \emph {et~al.}(2023)\citenamefont {Koch},
  \citenamefont {Mittleman}, \citenamefont {Ornik},\ and\ \citenamefont
  {Castro-Camus}}]{9}%
  \BibitemOpen
  \bibfield  {author} {\bibinfo {author} {\bibfnamefont {M.}~\bibnamefont
  {Koch}}, \bibinfo {author} {\bibfnamefont {D.~M.}\ \bibnamefont {Mittleman}},
  \bibinfo {author} {\bibfnamefont {J.}~\bibnamefont {Ornik}}, \ and\ \bibinfo
  {author} {\bibfnamefont {E.}~\bibnamefont {Castro-Camus}},\ }\bibfield
  {title} {\enquote {\bibinfo {title} {Terahertz time-domain spectroscopy},}\
  }\href@noop {} {\bibfield  {journal} {\bibinfo  {journal} {Nature Reviews
  Methods Primers}\ }\textbf {\bibinfo {volume} {3}},\ \bibinfo {pages} {48}
  (\bibinfo {year} {2023})}\BibitemShut {NoStop}%
\bibitem [{\citenamefont {Talebi}, \citenamefont {Guo},\ and\ \citenamefont
  {van Aken}(2018)}]{10}%
  \BibitemOpen
  \bibfield  {author} {\bibinfo {author} {\bibfnamefont {N.}~\bibnamefont
  {Talebi}}, \bibinfo {author} {\bibfnamefont {S.}~\bibnamefont {Guo}}, \ and\
  \bibinfo {author} {\bibfnamefont {P.~A.}\ \bibnamefont {van Aken}},\
  }\bibfield  {title} {\enquote {\bibinfo {title} {Theory and applications of
  toroidal moments in electrodynamics: their emergence, characteristics, and
  technological relevance},}\ }\href@noop {} {\bibfield  {journal} {\bibinfo
  {journal} {Nanophotonics}\ }\textbf {\bibinfo {volume} {7}},\ \bibinfo
  {pages} {93--110} (\bibinfo {year} {2018})}\BibitemShut {NoStop}%
\bibitem [{\citenamefont {Gupta}\ and\ \citenamefont {Singh}(2020)}]{11}%
  \BibitemOpen
  \bibfield  {author} {\bibinfo {author} {\bibfnamefont {M.}~\bibnamefont
  {Gupta}}\ and\ \bibinfo {author} {\bibfnamefont {R.}~\bibnamefont {Singh}},\
  }\bibfield  {title} {\enquote {\bibinfo {title} {Toroidal metasurfaces in a
  2d flatland},}\ }\href@noop {} {\bibfield  {journal} {\bibinfo  {journal}
  {Reviews in Physics}\ }\textbf {\bibinfo {volume} {5}},\ \bibinfo {pages}
  {100040} (\bibinfo {year} {2020})}\BibitemShut {NoStop}%
\bibitem [{\citenamefont {Ederer}\ and\ \citenamefont {Spaldin}(2007)}]{12}%
  \BibitemOpen
  \bibfield  {author} {\bibinfo {author} {\bibfnamefont {C.}~\bibnamefont
  {Ederer}}\ and\ \bibinfo {author} {\bibfnamefont {N.~A.}\ \bibnamefont
  {Spaldin}},\ }\bibfield  {title} {\enquote {\bibinfo {title} {Towards a
  microscopic theory of toroidal moments in bulk periodic crystals},}\
  }\href@noop {} {\bibfield  {journal} {\bibinfo  {journal} {Physical Review
  B}\ }\textbf {\bibinfo {volume} {76}},\ \bibinfo {pages} {214404} (\bibinfo
  {year} {2007})}\BibitemShut {NoStop}%
\bibitem [{\citenamefont {Marinov}\ \emph {et~al.}(2007)\citenamefont
  {Marinov}, \citenamefont {Boardman}, \citenamefont {Fedotov},\ and\
  \citenamefont {Zheludev}}]{13}%
  \BibitemOpen
  \bibfield  {author} {\bibinfo {author} {\bibfnamefont {K.}~\bibnamefont
  {Marinov}}, \bibinfo {author} {\bibfnamefont {A.}~\bibnamefont {Boardman}},
  \bibinfo {author} {\bibfnamefont {V.}~\bibnamefont {Fedotov}}, \ and\
  \bibinfo {author} {\bibfnamefont {N.}~\bibnamefont {Zheludev}},\ }\bibfield
  {title} {\enquote {\bibinfo {title} {Toroidal metamaterial},}\ }\href@noop {}
  {\bibfield  {journal} {\bibinfo  {journal} {New Journal of Physics}\ }\textbf
  {\bibinfo {volume} {9}},\ \bibinfo {pages} {324} (\bibinfo {year}
  {2007})}\BibitemShut {NoStop}%
\bibitem [{\citenamefont {Kaelberer}\ \emph {et~al.}(2010)\citenamefont
  {Kaelberer}, \citenamefont {Fedotov}, \citenamefont {Papasimakis},
  \citenamefont {Tsai},\ and\ \citenamefont {Zheludev}}]{14}%
  \BibitemOpen
  \bibfield  {author} {\bibinfo {author} {\bibfnamefont {T.}~\bibnamefont
  {Kaelberer}}, \bibinfo {author} {\bibfnamefont {V.}~\bibnamefont {Fedotov}},
  \bibinfo {author} {\bibfnamefont {N.}~\bibnamefont {Papasimakis}}, \bibinfo
  {author} {\bibfnamefont {D.}~\bibnamefont {Tsai}}, \ and\ \bibinfo {author}
  {\bibfnamefont {N.}~\bibnamefont {Zheludev}},\ }\bibfield  {title} {\enquote
  {\bibinfo {title} {Toroidal dipolar response in a metamaterial},}\
  }\href@noop {} {\bibfield  {journal} {\bibinfo  {journal} {Science}\ }\textbf
  {\bibinfo {volume} {330}},\ \bibinfo {pages} {1510--1512} (\bibinfo {year}
  {2010})}\BibitemShut {NoStop}%
\bibitem [{\citenamefont {Chen}, \citenamefont {Fan},\ and\ \citenamefont
  {Yan}(2020)}]{15}%
  \BibitemOpen
  \bibfield  {author} {\bibinfo {author} {\bibfnamefont {X.}~\bibnamefont
  {Chen}}, \bibinfo {author} {\bibfnamefont {W.}~\bibnamefont {Fan}}, \ and\
  \bibinfo {author} {\bibfnamefont {H.}~\bibnamefont {Yan}},\ }\bibfield
  {title} {\enquote {\bibinfo {title} {Toroidal dipole bound states in the
  continuum metasurfaces for terahertz nanofilm sensing},}\ }\href@noop {}
  {\bibfield  {journal} {\bibinfo  {journal} {Optics Express}\ }\textbf
  {\bibinfo {volume} {28}},\ \bibinfo {pages} {17102--17112} (\bibinfo {year}
  {2020})}\BibitemShut {NoStop}%
\bibitem [{\citenamefont {Savinov}, \citenamefont {Fedotov},\ and\
  \citenamefont {Zheludev}(2014)}]{16}%
  \BibitemOpen
  \bibfield  {author} {\bibinfo {author} {\bibfnamefont {V.}~\bibnamefont
  {Savinov}}, \bibinfo {author} {\bibfnamefont {V.}~\bibnamefont {Fedotov}}, \
  and\ \bibinfo {author} {\bibfnamefont {N.~I.}\ \bibnamefont {Zheludev}},\
  }\bibfield  {title} {\enquote {\bibinfo {title} {Toroidal dipolar excitation
  and macroscopic electromagnetic properties of metamaterials},}\ }\href@noop
  {} {\bibfield  {journal} {\bibinfo  {journal} {Physical Review B}\ }\textbf
  {\bibinfo {volume} {89}},\ \bibinfo {pages} {205112} (\bibinfo {year}
  {2014})}\BibitemShut {NoStop}%
\bibitem [{\citenamefont {Bhattacharya}, \citenamefont {Sarkar},\ and\
  \citenamefont {Kumar}(2021)}]{17}%
  \BibitemOpen
  \bibfield  {author} {\bibinfo {author} {\bibfnamefont {A.}~\bibnamefont
  {Bhattacharya}}, \bibinfo {author} {\bibfnamefont {R.}~\bibnamefont
  {Sarkar}}, \ and\ \bibinfo {author} {\bibfnamefont {G.}~\bibnamefont
  {Kumar}},\ }\bibfield  {title} {\enquote {\bibinfo {title} {Excitation of
  near field coupled dual toroidal resonances in a bilayer terahertz
  metamaterial configuration},}\ }\href@noop {} {\bibfield  {journal} {\bibinfo
   {journal} {Journal of Physics D: Applied Physics}\ }\textbf {\bibinfo
  {volume} {54}},\ \bibinfo {pages} {285102} (\bibinfo {year}
  {2021})}\BibitemShut {NoStop}%
\bibitem [{\citenamefont {Gupta}\ \emph {et~al.}(2017)\citenamefont {Gupta},
  \citenamefont {Srivastava}, \citenamefont {Manjappa},\ and\ \citenamefont
  {Singh}}]{18}%
  \BibitemOpen
  \bibfield  {author} {\bibinfo {author} {\bibfnamefont {M.}~\bibnamefont
  {Gupta}}, \bibinfo {author} {\bibfnamefont {Y.~K.}\ \bibnamefont
  {Srivastava}}, \bibinfo {author} {\bibfnamefont {M.}~\bibnamefont
  {Manjappa}}, \ and\ \bibinfo {author} {\bibfnamefont {R.}~\bibnamefont
  {Singh}},\ }\bibfield  {title} {\enquote {\bibinfo {title} {Sensing with
  toroidal metamaterial},}\ }\href@noop {} {\bibfield  {journal} {\bibinfo
  {journal} {Applied Physics Letters}\ }\textbf {\bibinfo {volume} {110}}
  (\bibinfo {year} {2017})}\BibitemShut {NoStop}%
\bibitem [{\citenamefont {Gupta}\ \emph {et~al.}(2016)\citenamefont {Gupta},
  \citenamefont {Savinov}, \citenamefont {Xu}, \citenamefont {Cong},
  \citenamefont {Dayal}, \citenamefont {Wang}, \citenamefont {Zhang},
  \citenamefont {Zheludev},\ and\ \citenamefont {Singh}}]{19}%
  \BibitemOpen
  \bibfield  {author} {\bibinfo {author} {\bibfnamefont {M.}~\bibnamefont
  {Gupta}}, \bibinfo {author} {\bibfnamefont {V.}~\bibnamefont {Savinov}},
  \bibinfo {author} {\bibfnamefont {N.}~\bibnamefont {Xu}}, \bibinfo {author}
  {\bibfnamefont {L.}~\bibnamefont {Cong}}, \bibinfo {author} {\bibfnamefont
  {G.}~\bibnamefont {Dayal}}, \bibinfo {author} {\bibfnamefont
  {S.}~\bibnamefont {Wang}}, \bibinfo {author} {\bibfnamefont {W.}~\bibnamefont
  {Zhang}}, \bibinfo {author} {\bibfnamefont {N.~I.}\ \bibnamefont {Zheludev}},
  \ and\ \bibinfo {author} {\bibfnamefont {R.}~\bibnamefont {Singh}},\
  }\bibfield  {title} {\enquote {\bibinfo {title} {Sharp toroidal resonances in
  planar terahertz metasurfaces},}\ }\href@noop {} {\bibfield  {journal}
  {\bibinfo  {journal} {Advanced Materials}\ }\textbf {\bibinfo {volume}
  {28}},\ \bibinfo {pages} {8206--8211} (\bibinfo {year} {2016})}\BibitemShut
  {NoStop}%
\bibitem [{\citenamefont {Ahmadivand}\ \emph {et~al.}(2020)\citenamefont
  {Ahmadivand}, \citenamefont {Gerislioglu}, \citenamefont {Ahuja},\ and\
  \citenamefont {Mishra}}]{20}%
  \BibitemOpen
  \bibfield  {author} {\bibinfo {author} {\bibfnamefont {A.}~\bibnamefont
  {Ahmadivand}}, \bibinfo {author} {\bibfnamefont {B.}~\bibnamefont
  {Gerislioglu}}, \bibinfo {author} {\bibfnamefont {R.}~\bibnamefont {Ahuja}},
  \ and\ \bibinfo {author} {\bibfnamefont {Y.~K.}\ \bibnamefont {Mishra}},\
  }\bibfield  {title} {\enquote {\bibinfo {title} {Terahertz plasmonics: The
  rise of toroidal metadevices towards immunobiosensings},}\ }\href@noop {}
  {\bibfield  {journal} {\bibinfo  {journal} {Materials Today}\ }\textbf
  {\bibinfo {volume} {32}},\ \bibinfo {pages} {108--130} (\bibinfo {year}
  {2020})}\BibitemShut {NoStop}%
\bibitem [{\citenamefont {Bhattacharya}, \citenamefont {Sarkar},\ and\
  \citenamefont {Kumar}(2022)}]{21}%
  \BibitemOpen
  \bibfield  {author} {\bibinfo {author} {\bibfnamefont {A.}~\bibnamefont
  {Bhattacharya}}, \bibinfo {author} {\bibfnamefont {R.}~\bibnamefont
  {Sarkar}}, \ and\ \bibinfo {author} {\bibfnamefont {G.}~\bibnamefont
  {Kumar}},\ }\bibfield  {title} {\enquote {\bibinfo {title} {Toroidal
  electromagnetically induced transparency based meta-surfaces and its
  applications},}\ }\href@noop {} {\bibfield  {journal} {\bibinfo  {journal}
  {Iscience}\ }\textbf {\bibinfo {volume} {25}},\ \bibinfo {pages} {103708}
  (\bibinfo {year} {2022})}\BibitemShut {NoStop}%
\bibitem [{\citenamefont {Bhattacharya}\ \emph {et~al.}(2021)\citenamefont
  {Bhattacharya}, \citenamefont {Sarkar}, \citenamefont {Sharma}, \citenamefont
  {Bhowmik}, \citenamefont {Ahmad},\ and\ \citenamefont {Kumar}}]{22}%
  \BibitemOpen
  \bibfield  {author} {\bibinfo {author} {\bibfnamefont {A.}~\bibnamefont
  {Bhattacharya}}, \bibinfo {author} {\bibfnamefont {R.}~\bibnamefont
  {Sarkar}}, \bibinfo {author} {\bibfnamefont {N.~K.}\ \bibnamefont {Sharma}},
  \bibinfo {author} {\bibfnamefont {B.~K.}\ \bibnamefont {Bhowmik}}, \bibinfo
  {author} {\bibfnamefont {A.}~\bibnamefont {Ahmad}}, \ and\ \bibinfo {author}
  {\bibfnamefont {G.}~\bibnamefont {Kumar}},\ }\bibfield  {title} {\enquote
  {\bibinfo {title} {Multiband transparency effect induced by toroidal
  excitation in a strongly coupled planar terahertz metamaterial},}\
  }\href@noop {} {\bibfield  {journal} {\bibinfo  {journal} {Scientific
  reports}\ }\textbf {\bibinfo {volume} {11}},\ \bibinfo {pages} {19186}
  (\bibinfo {year} {2021})}\BibitemShut {NoStop}%
\bibitem [{\citenamefont {Mallick}\ \emph {et~al.}(2023)\citenamefont
  {Mallick}, \citenamefont {Rane}, \citenamefont {Acharyya},\ and\
  \citenamefont {Chowdhury}}]{23}%
  \BibitemOpen
  \bibfield  {author} {\bibinfo {author} {\bibfnamefont {S.}~\bibnamefont
  {Mallick}}, \bibinfo {author} {\bibfnamefont {S.}~\bibnamefont {Rane}},
  \bibinfo {author} {\bibfnamefont {N.}~\bibnamefont {Acharyya}}, \ and\
  \bibinfo {author} {\bibfnamefont {D.~R.}\ \bibnamefont {Chowdhury}},\
  }\bibfield  {title} {\enquote {\bibinfo {title} {Accessing dual toroidal
  modes in terahertz plasmonic metasurfaces through polarization-sensitive
  resonance hybridization},}\ }\href@noop {} {\bibfield  {journal} {\bibinfo
  {journal} {New Journal of Physics}\ }\textbf {\bibinfo {volume} {25}},\
  \bibinfo {pages} {053016} (\bibinfo {year} {2023})}\BibitemShut {NoStop}%
\bibitem [{\citenamefont {Wang}\ \emph {et~al.}(2019)\citenamefont {Wang},
  \citenamefont {Wang}, \citenamefont {Zhao}, \citenamefont {Zhu},
  \citenamefont {Liu},\ and\ \citenamefont {Li}}]{232}%
  \BibitemOpen
  \bibfield  {author} {\bibinfo {author} {\bibfnamefont {S.}~\bibnamefont
  {Wang}}, \bibinfo {author} {\bibfnamefont {S.}~\bibnamefont {Wang}}, \bibinfo
  {author} {\bibfnamefont {X.}~\bibnamefont {Zhao}}, \bibinfo {author}
  {\bibfnamefont {J.}~\bibnamefont {Zhu}}, \bibinfo {author} {\bibfnamefont
  {S.}~\bibnamefont {Liu}}, \ and\ \bibinfo {author} {\bibfnamefont
  {Q.}~\bibnamefont {Li}},\ }\bibfield  {title} {\enquote {\bibinfo {title}
  {Dual toroidal dipole resonances in a planar terahertz flexible
  metasurfaces},}\ }\href@noop {} {\bibfield  {journal} {\bibinfo  {journal}
  {Materials Research Express}\ }\textbf {\bibinfo {volume} {6}},\ \bibinfo
  {pages} {115803} (\bibinfo {year} {2019})}\BibitemShut {NoStop}%
\bibitem [{\citenamefont {Basharin}\ \emph {et~al.}(2017)\citenamefont
  {Basharin}, \citenamefont {Chuguevsky}, \citenamefont {Volsky}, \citenamefont
  {Kafesaki},\ and\ \citenamefont {Economou}}]{24}%
  \BibitemOpen
  \bibfield  {author} {\bibinfo {author} {\bibfnamefont {A.~A.}\ \bibnamefont
  {Basharin}}, \bibinfo {author} {\bibfnamefont {V.}~\bibnamefont
  {Chuguevsky}}, \bibinfo {author} {\bibfnamefont {N.}~\bibnamefont {Volsky}},
  \bibinfo {author} {\bibfnamefont {M.}~\bibnamefont {Kafesaki}}, \ and\
  \bibinfo {author} {\bibfnamefont {E.~N.}\ \bibnamefont {Economou}},\
  }\bibfield  {title} {\enquote {\bibinfo {title} {Extremely high q-factor
  metamaterials due to anapole excitation},}\ }\href@noop {} {\bibfield
  {journal} {\bibinfo  {journal} {Physical Review B}\ }\textbf {\bibinfo
  {volume} {95}},\ \bibinfo {pages} {035104} (\bibinfo {year}
  {2017})}\BibitemShut {NoStop}%
\bibitem [{\citenamefont {Fan}\ \emph {et~al.}(2018)\citenamefont {Fan},
  \citenamefont {Zhang}, \citenamefont {Shen}, \citenamefont {Fu},
  \citenamefont {Wei}, \citenamefont {Li},\ and\ \citenamefont
  {Soukoulis}}]{25}%
  \BibitemOpen
  \bibfield  {author} {\bibinfo {author} {\bibfnamefont {Y.}~\bibnamefont
  {Fan}}, \bibinfo {author} {\bibfnamefont {F.}~\bibnamefont {Zhang}}, \bibinfo
  {author} {\bibfnamefont {N.-H.}\ \bibnamefont {Shen}}, \bibinfo {author}
  {\bibfnamefont {Q.}~\bibnamefont {Fu}}, \bibinfo {author} {\bibfnamefont
  {Z.}~\bibnamefont {Wei}}, \bibinfo {author} {\bibfnamefont {H.}~\bibnamefont
  {Li}}, \ and\ \bibinfo {author} {\bibfnamefont {C.~M.}\ \bibnamefont
  {Soukoulis}},\ }\bibfield  {title} {\enquote {\bibinfo {title} {Achieving a
  high-q response in metamaterials by manipulating the toroidal excitations},}\
  }\href@noop {} {\bibfield  {journal} {\bibinfo  {journal} {Physical Review
  A}\ }\textbf {\bibinfo {volume} {97}},\ \bibinfo {pages} {033816} (\bibinfo
  {year} {2018})}\BibitemShut {NoStop}%
\bibitem [{\citenamefont {Lou}\ \emph {et~al.}(2021)\citenamefont {Lou},
  \citenamefont {Yang}, \citenamefont {Liang}, \citenamefont {Yu},
  \citenamefont {Zhang}, \citenamefont {Zhang}, \citenamefont {Li},
  \citenamefont {Fan}, \citenamefont {Zhang}, \citenamefont {Wang} \emph
  {et~al.}}]{26}%
  \BibitemOpen
  \bibfield  {author} {\bibinfo {author} {\bibfnamefont {J.}~\bibnamefont
  {Lou}}, \bibinfo {author} {\bibfnamefont {R.}~\bibnamefont {Yang}}, \bibinfo
  {author} {\bibfnamefont {J.}~\bibnamefont {Liang}}, \bibinfo {author}
  {\bibfnamefont {Y.}~\bibnamefont {Yu}}, \bibinfo {author} {\bibfnamefont
  {L.}~\bibnamefont {Zhang}}, \bibinfo {author} {\bibfnamefont
  {C.}~\bibnamefont {Zhang}}, \bibinfo {author} {\bibfnamefont
  {T.}~\bibnamefont {Li}}, \bibinfo {author} {\bibfnamefont {Y.}~\bibnamefont
  {Fan}}, \bibinfo {author} {\bibfnamefont {F.}~\bibnamefont {Zhang}}, \bibinfo
  {author} {\bibfnamefont {G.}~\bibnamefont {Wang}},  \emph {et~al.},\
  }\bibfield  {title} {\enquote {\bibinfo {title} {Dual-sensitivity terahertz
  metasensor based on lattice--toroidal-coupled resonance},}\ }\href@noop {}
  {\bibfield  {journal} {\bibinfo  {journal} {Advanced Photonics Research}\
  }\textbf {\bibinfo {volume} {2}},\ \bibinfo {pages} {2000175} (\bibinfo
  {year} {2021})}\BibitemShut {NoStop}%
\bibitem [{\citenamefont {Gupta}\ and\ \citenamefont {Singh}(2016)}]{27}%
  \BibitemOpen
  \bibfield  {author} {\bibinfo {author} {\bibfnamefont {M.}~\bibnamefont
  {Gupta}}\ and\ \bibinfo {author} {\bibfnamefont {R.}~\bibnamefont {Singh}},\
  }\bibfield  {title} {\enquote {\bibinfo {title} {Toroidal versus fano
  resonances in high q planar thz metamaterials},}\ }\href@noop {} {\bibfield
  {journal} {\bibinfo  {journal} {Advanced Optical Materials}\ }\textbf
  {\bibinfo {volume} {4}},\ \bibinfo {pages} {2119--2125} (\bibinfo {year}
  {2016})}\BibitemShut {NoStop}%
\bibitem [{\citenamefont {Fushimi}\ and\ \citenamefont {Tanabe}(2014)}]{30}%
  \BibitemOpen
  \bibfield  {author} {\bibinfo {author} {\bibfnamefont {A.}~\bibnamefont
  {Fushimi}}\ and\ \bibinfo {author} {\bibfnamefont {T.}~\bibnamefont
  {Tanabe}},\ }\bibfield  {title} {\enquote {\bibinfo {title} {All-optical
  logic gate operating with single wavelength},}\ }\href@noop {} {\bibfield
  {journal} {\bibinfo  {journal} {Optics express}\ }\textbf {\bibinfo {volume}
  {22}},\ \bibinfo {pages} {4466--4479} (\bibinfo {year} {2014})}\BibitemShut
  {NoStop}%
\bibitem [{\citenamefont {Jiao}\ \emph {et~al.}(2022)\citenamefont {Jiao},
  \citenamefont {Liu}, \citenamefont {Zhang}, \citenamefont {Yu}, \citenamefont
  {Zuo}, \citenamefont {Zhang}, \citenamefont {Zhao}, \citenamefont {Lin},\
  and\ \citenamefont {Shao}}]{31}%
  \BibitemOpen
  \bibfield  {author} {\bibinfo {author} {\bibfnamefont {S.}~\bibnamefont
  {Jiao}}, \bibinfo {author} {\bibfnamefont {J.}~\bibnamefont {Liu}}, \bibinfo
  {author} {\bibfnamefont {L.}~\bibnamefont {Zhang}}, \bibinfo {author}
  {\bibfnamefont {F.}~\bibnamefont {Yu}}, \bibinfo {author} {\bibfnamefont
  {G.}~\bibnamefont {Zuo}}, \bibinfo {author} {\bibfnamefont {J.}~\bibnamefont
  {Zhang}}, \bibinfo {author} {\bibfnamefont {F.}~\bibnamefont {Zhao}},
  \bibinfo {author} {\bibfnamefont {W.}~\bibnamefont {Lin}}, \ and\ \bibinfo
  {author} {\bibfnamefont {L.}~\bibnamefont {Shao}},\ }\bibfield  {title}
  {\enquote {\bibinfo {title} {All-optical logic gate computing for high-speed
  parallel information processing},}\ }\href@noop {} {\bibfield  {journal}
  {\bibinfo  {journal} {Opto-Electronic Science}\ }\textbf {\bibinfo {volume}
  {1}},\ \bibinfo {pages} {220010--1} (\bibinfo {year} {2022})}\BibitemShut
  {NoStop}%
\bibitem [{\citenamefont {Manjappa}\ \emph {et~al.}(2018)\citenamefont
  {Manjappa}, \citenamefont {Pitchappa}, \citenamefont {Singh}, \citenamefont
  {Wang}, \citenamefont {Zheludev}, \citenamefont {Lee},\ and\ \citenamefont
  {Singh}}]{32}%
  \BibitemOpen
  \bibfield  {author} {\bibinfo {author} {\bibfnamefont {M.}~\bibnamefont
  {Manjappa}}, \bibinfo {author} {\bibfnamefont {P.}~\bibnamefont {Pitchappa}},
  \bibinfo {author} {\bibfnamefont {N.}~\bibnamefont {Singh}}, \bibinfo
  {author} {\bibfnamefont {N.}~\bibnamefont {Wang}}, \bibinfo {author}
  {\bibfnamefont {N.~I.}\ \bibnamefont {Zheludev}}, \bibinfo {author}
  {\bibfnamefont {C.}~\bibnamefont {Lee}}, \ and\ \bibinfo {author}
  {\bibfnamefont {R.}~\bibnamefont {Singh}},\ }\bibfield  {title} {\enquote
  {\bibinfo {title} {Reconfigurable mems fano metasurfaces with
  multiple-input--output states for logic operations at terahertz
  frequencies},}\ }\href@noop {} {\bibfield  {journal} {\bibinfo  {journal}
  {Nature communications}\ }\textbf {\bibinfo {volume} {9}},\ \bibinfo {pages}
  {4056} (\bibinfo {year} {2018})}\BibitemShut {NoStop}%
\bibitem [{\citenamefont {Wang}\ \emph {et~al.}(2022)\citenamefont {Wang},
  \citenamefont {Zhang}, \citenamefont {Qiu}, \citenamefont {Wang},
  \citenamefont {Yang}, \citenamefont {Li}, \citenamefont {Hu}, \citenamefont
  {Li}, \citenamefont {Yan}, \citenamefont {Yao} \emph {et~al.}}]{33}%
  \BibitemOpen
  \bibfield  {author} {\bibinfo {author} {\bibfnamefont {Z.}~\bibnamefont
  {Wang}}, \bibinfo {author} {\bibfnamefont {Z.}~\bibnamefont {Zhang}},
  \bibinfo {author} {\bibfnamefont {F.}~\bibnamefont {Qiu}}, \bibinfo {author}
  {\bibfnamefont {M.}~\bibnamefont {Wang}}, \bibinfo {author} {\bibfnamefont
  {W.}~\bibnamefont {Yang}}, \bibinfo {author} {\bibfnamefont {Z.}~\bibnamefont
  {Li}}, \bibinfo {author} {\bibfnamefont {X.}~\bibnamefont {Hu}}, \bibinfo
  {author} {\bibfnamefont {Y.}~\bibnamefont {Li}}, \bibinfo {author}
  {\bibfnamefont {X.}~\bibnamefont {Yan}}, \bibinfo {author} {\bibfnamefont
  {H.}~\bibnamefont {Yao}},  \emph {et~al.},\ }\bibfield  {title} {\enquote
  {\bibinfo {title} {Design of an all-optical multi-logic operation-integrated
  metamaterial-based terahertz logic gate},}\ }\href@noop {} {\bibfield
  {journal} {\bibinfo  {journal} {Optics Express}\ }\textbf {\bibinfo {volume}
  {30}},\ \bibinfo {pages} {40401--40412} (\bibinfo {year} {2022})}\BibitemShut
  {NoStop}%
\bibitem [{\citenamefont {Yuan}\ \emph {et~al.}(2020)\citenamefont {Yuan},
  \citenamefont {Wang}, \citenamefont {Li}, \citenamefont {Xu}, \citenamefont
  {Xu}, \citenamefont {Zhang}, \citenamefont {Zhang}, \citenamefont {Han},\
  and\ \citenamefont {Zhang}}]{34}%
  \BibitemOpen
  \bibfield  {author} {\bibinfo {author} {\bibfnamefont {M.}~\bibnamefont
  {Yuan}}, \bibinfo {author} {\bibfnamefont {Q.}~\bibnamefont {Wang}}, \bibinfo
  {author} {\bibfnamefont {Y.}~\bibnamefont {Li}}, \bibinfo {author}
  {\bibfnamefont {Y.}~\bibnamefont {Xu}}, \bibinfo {author} {\bibfnamefont
  {Q.}~\bibnamefont {Xu}}, \bibinfo {author} {\bibfnamefont {X.}~\bibnamefont
  {Zhang}}, \bibinfo {author} {\bibfnamefont {X.}~\bibnamefont {Zhang}},
  \bibinfo {author} {\bibfnamefont {J.}~\bibnamefont {Han}}, \ and\ \bibinfo
  {author} {\bibfnamefont {W.}~\bibnamefont {Zhang}},\ }\bibfield  {title}
  {\enquote {\bibinfo {title} {Terahertz spoof surface plasmonic logic
  gates},}\ }\href@noop {} {\bibfield  {journal} {\bibinfo  {journal}
  {Iscience}\ }\textbf {\bibinfo {volume} {23}} (\bibinfo {year}
  {2020})}\BibitemShut {NoStop}%
\bibitem [{\citenamefont {Xu}, \citenamefont {Xu},\ and\ \citenamefont
  {Lin}(2022)}]{35}%
  \BibitemOpen
  \bibfield  {author} {\bibinfo {author} {\bibfnamefont {R.}~\bibnamefont
  {Xu}}, \bibinfo {author} {\bibfnamefont {X.}~\bibnamefont {Xu}}, \ and\
  \bibinfo {author} {\bibfnamefont {Y.-S.}\ \bibnamefont {Lin}},\ }\bibfield
  {title} {\enquote {\bibinfo {title} {Electrothermally tunable terahertz
  cross-shaped metamaterial for opto-logic operation characteristics},}\
  }\href@noop {} {\bibfield  {journal} {\bibinfo  {journal} {Iscience}\
  }\textbf {\bibinfo {volume} {25}} (\bibinfo {year} {2022})}\BibitemShut
  {NoStop}%
\bibitem [{\citenamefont {Meymand}, \citenamefont {Soleymani},\ and\
  \citenamefont {Granpayeh}(2020)}]{36}%
  \BibitemOpen
  \bibfield  {author} {\bibinfo {author} {\bibfnamefont {R.~E.}\ \bibnamefont
  {Meymand}}, \bibinfo {author} {\bibfnamefont {A.}~\bibnamefont {Soleymani}},
  \ and\ \bibinfo {author} {\bibfnamefont {N.}~\bibnamefont {Granpayeh}},\
  }\bibfield  {title} {\enquote {\bibinfo {title} {All-optical and, or, and xor
  logic gates based on coherent perfect absorption in graphene-based
  metasurface at terahertz region},}\ }\href@noop {} {\bibfield  {journal}
  {\bibinfo  {journal} {Optics Communications}\ }\textbf {\bibinfo {volume}
  {458}},\ \bibinfo {pages} {124772} (\bibinfo {year} {2020})}\BibitemShut
  {NoStop}%
\bibitem [{\citenamefont {Xie}\ \emph {et~al.}(2022)\citenamefont {Xie},
  \citenamefont {Yin}, \citenamefont {Song}, \citenamefont {Zhu}, \citenamefont
  {Chai}, \citenamefont {Liu},\ and\ \citenamefont {Ye}}]{37}%
  \BibitemOpen
  \bibfield  {author} {\bibinfo {author} {\bibfnamefont {Y.}~\bibnamefont
  {Xie}}, \bibinfo {author} {\bibfnamefont {Y.}~\bibnamefont {Yin}}, \bibinfo
  {author} {\bibfnamefont {T.}~\bibnamefont {Song}}, \bibinfo {author}
  {\bibfnamefont {Y.}~\bibnamefont {Zhu}}, \bibinfo {author} {\bibfnamefont
  {J.}~\bibnamefont {Chai}}, \bibinfo {author} {\bibfnamefont {B.}~\bibnamefont
  {Liu}}, \ and\ \bibinfo {author} {\bibfnamefont {Y.}~\bibnamefont {Ye}},\
  }\bibfield  {title} {\enquote {\bibinfo {title} {The design and simulation of
  a multifunctional logic device based on plasmon-induced transparency using
  two semicircular resonators},}\ }\href@noop {} {\bibfield  {journal}
  {\bibinfo  {journal} {Optik}\ }\textbf {\bibinfo {volume} {255}},\ \bibinfo
  {pages} {168684} (\bibinfo {year} {2022})}\BibitemShut {NoStop}%
\bibitem [{\citenamefont {Bhattacharya}\ \emph {et~al.}(2023)\citenamefont
  {Bhattacharya}, \citenamefont {Chouhan}, \citenamefont {Bhowmik},
  \citenamefont {Singh},\ and\ \citenamefont {Kumar}}]{38}%
  \BibitemOpen
  \bibfield  {author} {\bibinfo {author} {\bibfnamefont {A.}~\bibnamefont
  {Bhattacharya}}, \bibinfo {author} {\bibfnamefont {B.~S.}\ \bibnamefont
  {Chouhan}}, \bibinfo {author} {\bibfnamefont {B.~K.}\ \bibnamefont
  {Bhowmik}}, \bibinfo {author} {\bibfnamefont {R.}~\bibnamefont {Singh}}, \
  and\ \bibinfo {author} {\bibfnamefont {G.}~\bibnamefont {Kumar}},\ }\bibfield
   {title} {\enquote {\bibinfo {title} {Polarization independent
  lattice-coupled terahertz toroidal excitations},}\ }\href@noop {} {\bibfield
  {journal} {\bibinfo  {journal} {Journal of Physics D: Applied Physics}\
  }\textbf {\bibinfo {volume} {56}},\ \bibinfo {pages} {415101} (\bibinfo
  {year} {2023})}\BibitemShut {NoStop}%
\bibitem [{\citenamefont {Sarkar}\ \emph {et~al.}(2022)\citenamefont {Sarkar},
  \citenamefont {Bhattacharya}, \citenamefont {Punjal}, \citenamefont
  {Prabhu},\ and\ \citenamefont {Kumar}}]{28}%
  \BibitemOpen
  \bibfield  {author} {\bibinfo {author} {\bibfnamefont {R.}~\bibnamefont
  {Sarkar}}, \bibinfo {author} {\bibfnamefont {A.}~\bibnamefont
  {Bhattacharya}}, \bibinfo {author} {\bibfnamefont {A.}~\bibnamefont
  {Punjal}}, \bibinfo {author} {\bibfnamefont {S.~S.}\ \bibnamefont {Prabhu}},
  \ and\ \bibinfo {author} {\bibfnamefont {G.}~\bibnamefont {Kumar}},\
  }\bibfield  {title} {\enquote {\bibinfo {title} {Broadband terahertz
  polarization conversion using a planar toroidal metamaterial},}\ }\href@noop
  {} {\bibfield  {journal} {\bibinfo  {journal} {Journal of Applied Physics}\
  }\textbf {\bibinfo {volume} {132}} (\bibinfo {year} {2022})}\BibitemShut
  {NoStop}%
\bibitem [{\citenamefont {Bhattacharya}\ \emph {et~al.}(2020)\citenamefont
  {Bhattacharya}, \citenamefont {Devi}, \citenamefont {Nguyen},\ and\
  \citenamefont {Kumar}}]{29}%
  \BibitemOpen
  \bibfield  {author} {\bibinfo {author} {\bibfnamefont {A.}~\bibnamefont
  {Bhattacharya}}, \bibinfo {author} {\bibfnamefont {K.~M.}\ \bibnamefont
  {Devi}}, \bibinfo {author} {\bibfnamefont {T.}~\bibnamefont {Nguyen}}, \ and\
  \bibinfo {author} {\bibfnamefont {G.}~\bibnamefont {Kumar}},\ }\bibfield
  {title} {\enquote {\bibinfo {title} {Actively tunable toroidal excitations in
  graphene based terahertz metamaterials},}\ }\href@noop {} {\bibfield
  {journal} {\bibinfo  {journal} {Optics Communications}\ }\textbf {\bibinfo
  {volume} {459}},\ \bibinfo {pages} {124919} (\bibinfo {year}
  {2020})}\BibitemShut {NoStop}%
\end{thebibliography}%

\end{document}